%% file: main.tex
\documentclass{article}

\usepackage[preprint]{template}

\usepackage[utf8]{inputenc} 
\usepackage[T1]{fontenc}    
\usepackage{newtxtext}      
\usepackage{hyperref}       
\usepackage{url}            
\usepackage{booktabs}       
\usepackage{amsfonts}       
\usepackage{nicefrac}       
\usepackage{microtype}      
\usepackage{xcolor}         
\usepackage{xspace}
\usepackage[capitalize,noabbrev]{cleveref}
\usepackage{subfigure}
\usepackage{tcolorbox}
\usepackage{pifont}
\usepackage{amssymb}
\usepackage{bbding}
\usepackage{soul}
\usepackage{caption}
\usepackage{subcaption}
\usepackage{graphicx}
\usepackage{float}
\usepackage{wrapfig}
\usepackage{lipsum} 
\usepackage{algorithm}
\usepackage{algorithmic}

\newcommand{\name}{\textsc{Teraio}\xspace}

\title{Cost-Efficient LLM Training with Lifetime-Aware Tensor Offloading via GPUDirect Storage}

\author{
  Ziqi Yuan$^{1}$ \\
  \texttt{ziqiy6@illinois.edu} \\
  \And
  Haoyang Zhang$^{1}$ \\
  \texttt{zhang402@illinois.edu} \\
  \And
  Yirui Eric Zhou$^{1}$ \\
  \texttt{yiruiz2@illinois.edu} \\
  \AND
  Apoorve Mohan$^{2}$ \\
  \texttt{apoorve.mohan@ibm.com} \\
  \And
  I-Hsin Chung$^{2}$ \\
  \texttt{ihchung@us.ibm.com} \\
  \And
  Seetharami Seelam$^{2}$ \\
  \texttt{sseelam@us.ibm.com} \\
  \AND
  Jian Huang$^{1}$ \\
  \texttt{jianh@illinois.edu} \\
  \AND
  $^{1}$University of Illinois Urbana-Champaign\hspace{3em}$^{2}$IBM Research
}

\begin{document}
\newcommand{\code}[1]{\texttt{#1}}

\crefformat{section}{\S#2#1#3}
\crefformat{subsection}{\S#2#1#3}
\crefformat{subsubsection}{\S#2#1#3}
\crefname{figure}{Figure}{Figures}
\crefname{table}{Table}{Tables}
\crefname{algorithm}{Algorithm}{Algorithms}
\crefname{theorem}{CAP}{Capabilities}

\maketitle
\author{}

\begin{abstract}
We present the design and implementation of a new lifetime-aware tensor offloading framework for GPU memory expansion using low-cost PCIe-based solid-state drives (SSDs). Our framework, \name{}, is developed explicitly for large language model (LLM) training with multiple GPUs and multiple SSDs. Its design is driven by our observation that the active tensors take only a small fraction {(1.7\% on average)} of allocated GPU memory in each LLM training iteration, the inactive tensors are usually large and will not be used for a long period of time, creating ample opportunities for offloading/prefetching tensors to/from slow SSDs without stalling the GPU training process. \name{} accurately estimates the lifetime (active period of time in GPU memory) of each tensor with the profiling of the first few iterations in the training process. With the tensor lifetime analysis, \name{} will generate an optimized tensor offloading/prefetching plan and integrate it into the compiled LLM program via PyTorch. \name{} has a runtime tensor migration engine to execute the offloading/prefetching plan via GPUDirect storage, which allows direct tensor migration between GPUs and SSDs for alleviating the CPU bottleneck and maximizing the SSD bandwidth utilization. In comparison with state-of-the-art studies such as ZeRO-Offload and ZeRO-Infinity, we show that \name{} improves the training performance of various LLMs by {1.47}$\times$ on average, and achieves {80.7}\% of the ideal performance assuming unlimited GPU memory. 

\end{abstract}

\input{intro}

\input{characterization}

\input{design}
\input{eval}
\input{ending}
\bibliographystyle{plain}

\input{main.bbl}

\end{document}

%% file: intro.tex
\section{Introduction}

\vspace{-1ex}

Large language models (LLMs) have been widely employed
in various application domains~\cite{paper:diffusion_trans,paper:netllm, paper:agent_for_cloud}. 
However, training LLMs is still a well-known challenging problem, as its memory demand is increasing at a much faster speed than the scaling speed of GPU memory~\citep{paper:scaling_laws, paper:large_batch_training}. Scaling the GPU memory capacity is both technically challenging and prohibitively expensive. Due to space constraints and DRAM scaling issues~\citep{paper:memorywall, paper:checkmate}, it is hard to scale up the GPU memory capacity on each server machine. Scaling out LLM training with a cluster of GPU servers can increase the aggregated memory capacity, however, it inevitably increases operational cost, placing a major barrier for researchers and developers to entry into cutting-edge LLM development. 


To overcome the GPU memory wall, prior studies proposed expanding GPU memory with external memory devices.  One common approach is to allow tensor offloading from GPU memory to host memory~\cite{paper:deepum, paper:zero-offload, paper:swapadvisor, paper:superneurons, paper:l2l, paper:vdnn, paper:stronghold, paper:tflms}. However, due to the fundamental DRAM scalability challenge, such an approach is still limited by the host memory. Recent studies have extended tensor offloading to PCIe-based SSDs that offer larger capacity at a much lower cost~\cite{paper:flashneuron, paper:zero-infinity, paper:smart-infinity, paper:fuyou, paper:tba, paper:flashgpu, paper:angel-ptm, paper:zng}. 
But due to the incapability of efficiently utilizing the SSD bandwidth and hiding the slow SSD accesses, these offloading solutions still deliver suboptimal performance. For instance, ZeRO-infinity~\cite{paper:zero-infinity} enabled the offloading of tensors to SSDs at the granularity of deep neural network (DNN) layers, its coarse-grained offloading/prefetching scheme wastes not only the limited SSD bandwidth but also the precious GPU memory space, leaving the solution less attractive in practice.

Ideally, we wish to expand GPU memory with low-cost SSDs while achieving similar training performance as the ideal case assuming GPUs have unlimited on-board memory. To this end, we present a new tensor offloading framework -- \name{}, which enables fine-grained offloading of tensors in an accurate fashion based on their activity patterns in GPUs, for best utilizing both SSD bandwidth and GPU memory. Our study of tensor activity patterns ($\S$\ref{sec:study}) in LLM training shows 
(1) the active tensors, which are used in the current kernel during LLM training, consume only a small portion (1.7\% on average) of requested GPU memory in total; (2) many inactive tensors are large and occupy a substantial GPU memory space in each training iteration; but (3) these inactive tensors are not used in the training for a long period of time, depending on the computation intensity of LLMs. 

With these insights, \name{} develops a lifetime-aware tensor offloading mechanism following three design principles: (1) offloading large inactive tensors to SSDs can save precious GPU memory and maximize SSD bandwidth utilization during the training process; (2) the distribution of their inactive periods of time will help \name{} decide which inactive tensor should be offloaded at what time, and similarly, which tensor should be prefetched at what time; (3) precisely scheduling tensor offloading and prefetching in consideration of the available SSD bandwidth will help \name{} effectively overlap the tensor movement with GPU computation. Our roofline model analysis ($\S$\ref{subsec:bandwidth}) shows that, given each GPU connected to multiple commodity SSDs today, the aggregated storage I/O bandwidth is sufficient to meet the tensor migration requirement without hurting the GPU training process. 

To fulfill the design principles discussed above, we develop \name{} with three major components: (1) a tensor lifetime profiler that can extract tensor activity patterns (e.g., tensor size and lifetime) in advance with the assistance of deep learning compilers such as PyTorch, (2) a lifetime-aware tensor migration algorithm that can generate optimal tensor offloading/prefetching plans based on the learned tensor activity patterns, and (3) a tensor migration engine that will execute the generated offloading/prefetching plans with efficient direct data transfer between GPUs, host memory, and SSDs. We present each of these components as follows. 



\noindent
\textbf{An open-source tensor lifetime profiler.} 
\name{} conducts the profiling of the tensor size and lifetime distributions by running the first few iterations of LLM training on the target GPU setting. As the computation and dataflow patterns of each iteration are almost the same, the profiling results can accurately represent the generic patterns of the entire LLM training process.  
To track the metadata information of each tensor, we instrument the automatic operator generator in PyTorch rather than intrusively instrument the source code of each generated operator. Therefore, the proifler requires minimal code modifications to PyTorch. As the execution of LLM on GPUs has highly predictable dataflow patterns, \name{} uses the execution time of GPU kernels to accurately estimate the tensor lifetime (i.e., the length of the active and inactive period of time). With the knowledge of tensor activity patterns, \name{} will create the tensor offloading/prefetching plans in advance.




\noindent
\textbf{Lifetime-aware tensor migration algorithm.}
\name{} prioritizes offloading large tensors with long inactive period of time to SSDs, for fully utilizing the available storage I/O bandwidth. For tensors that have short inactive periods of time, \name{} will make the best effort to retain them in GPU memory, for avoiding unnecessary migration overhead. As host memory and SSD offer different capacities, bandwidths, and costs, \name{} prefers to offload tensors to SSDs for taking advantage of their large capacity and low cost. However, when the SSD bandwidth is saturated at runtime, \name{} will use the host memory as the offloading destination. Given an execution plan for each LLM training iteration, \name{} will iteratively search for the best offloading candidate based on the tensor size and lifetime, until the required GPU memory is below the capacity limit. After that, \name{} will generate an optimized tensor migration plan by adding corresponding offloading and prefetching instructions in the compiled LLM training program.

\noindent
\textbf{Tensor migration engine using GPUDirect storage.}
Following the tensor migration plan integrated into the compiled LLM training program, the tenor migration engine of \name{} will offload/prefetch tensors to/from SSDs or host memory at runtime. 
For the tensor migration between GPU memory and SSD, \name{} uses GPUDirect storage to enable direct data transfer between GPUs and SSDs, therefore, it can bypass the host CPU to alleviate the scalability bottleneck and maximize SSD bandwidth utilization. 
When the available SSD bandwidth is insufficient to support tensor offloading and prefetching, \name{} will migrate tensors to the host memory. 
To track the latest locations of tensors (GPU memory, host memory, or SSDs), \name{} indexes tensors with their identification numbers using hash maps. 

We implement the core components of \name{} based on PyTorch. Therefore, \name{} does not require any code modifications to LLM training programs. To evaluate the efficiency of \name{}, we train a set of Llama and Granite models with different batch sizes and sequence lengths using TorchTitan~\cite{paper:torchtitan} on a GPU server that has two NVIDIA H100 GPUs and eight PCIe-based SSDs. In comparison with state-of-the-art offloading solutions ZeRO-Offload~\cite{paper:zero-offload} and ZeRO-Infinity~\cite{paper:zero-infinity}, \name{} improves the training performance by 1.47$\times$ on average, achieves 80.7\% of the ideal performance assuming unlimited GPU on-board memory, and delivers 1.45$\times$ improvement on cost efficiency for LLM training. In summary, we make the following contributions. 


\begin{itemize}
    \item We conduct a quantitative characterization study of tensor memory usage when training different LLMs on multiple GPUs, and show that the high compute intensity of modern LLMs provide rich opportunities for tensor offloading.
    \item We develop a lightweight tensor lifetime profiler based on PyTorch, which can learn tensor activity patterns for multi-GPU LLM training. 
    \item We design a lifetime-aware tensor migration planning algorithm that optimizes offloading/prefetching decisions based on tensor activity patterns, GPU memory capacity, and the available migration bandwidth. 
    \item We implement a transparent tensor migration engine that enables direct data transfer between GPU and SSDs, alleviating the scalability bottleneck on the host.  
    \item We conduct a thorough evaluation of \name{} with 
    the training of various LLMs, demonstrating significant improvement on training performance and cost efficiency, compared to state-of-the-art offloading solutions. 
\end{itemize}

%% file: characterization.tex
\section{Characterization Study of Tensor Activity Patterns in LLM Training}
\label{sec:study}

In this section, we present our characterization study of tensor activity patterns in LLM training. To facilitate our study, we utilize our tensor lifetime profiler, which will be discussed in $\S$\ref{sec:profiling_tool}, to analyze the distributions of tensor sizes and lifetimes during the LLM training. We use {two NVIDIA H100 GPUs and 2-stage \texttt{1f1b}} pipeline parallelism \cite{paper:pipedream} in our experiments. Our study covers a variety of LLMs with different architectures, including decoder-only models such as Llama3-8B, Llama3-70B \cite{paper:llama}, and GPT2-40B \cite{gpt2}, as well as encoder-decoder models like T5-11B \cite{website:t5}. For models that require more memory than GPU memory capacity, we offload tensors not needed by the current kernel to SSDs. We summarize our study results as follows.

\begin{figure*}[!ht]
    \centering
    \vspace{-2ex}
    \begin{minipage}{0.96\linewidth}
        \centering
        \includegraphics[width=\linewidth]{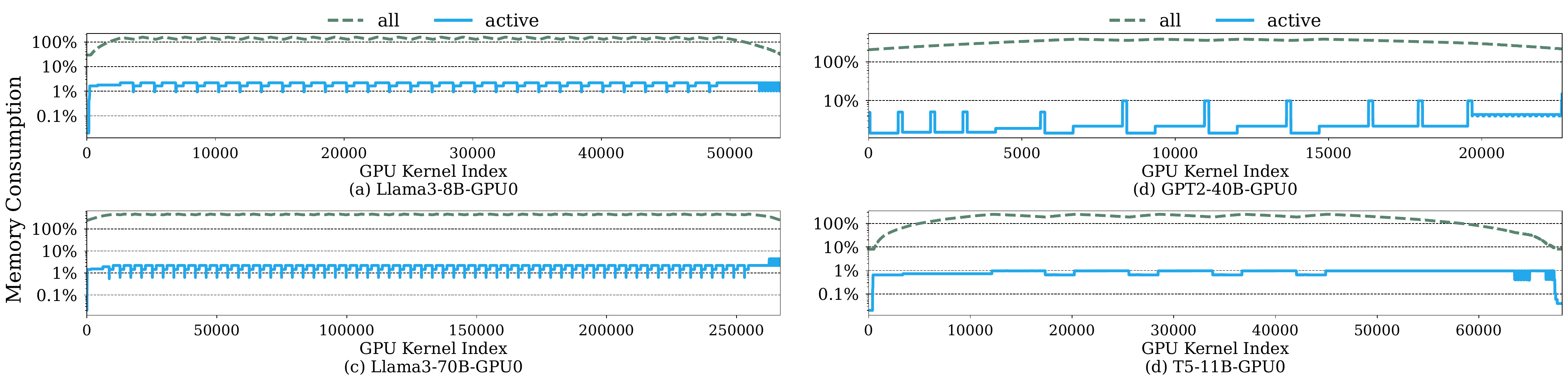}
        \vspace{-3ex}
        \caption*{(a) Memory consumption of different LLMs on each GPU.}
    \end{minipage}
    \hfill
    \begin{minipage}{0.96\linewidth}
        \centering
        \includegraphics[width=\linewidth]{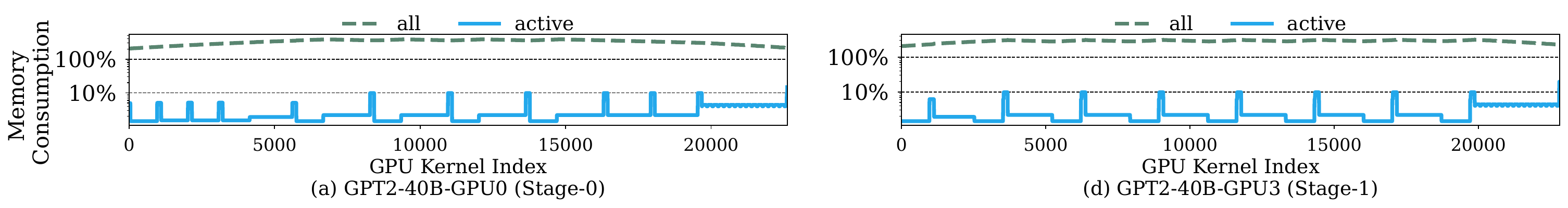}
           \vspace{-3ex}
        \caption*{(b) Memory consumption of different stages in the pipeline parallelism.}
    \end{minipage}
    \vspace{-1ex}
    \caption{Memory consumption of all and active tensors (w.r.t. the GPU memory capacity) in one iteration of the parallel training program. Logarithmic scale is used in the presentation.}
    \vspace{-2ex}
    \label{fig:tensor_mem_consumption}
\end{figure*}

\subsection{Rich Opportunities for Tensor Migration}

\noindent
\textbf{Small memory requirement of active tensors.}
We first study the memory demand and usage of tensors in each training iteration. We define the tensors that are currently used by a running GPU kernel as \textbf{\textit{active tensors}}.
We present the memory consumption of active tensors used by each GPU kernel in \cref{fig:tensor_mem_consumption}.
Figure~\ref{fig:tensor_mem_consumption} (a) shows the memory consumption of tensors in different models, Figure~\ref{fig:tensor_mem_consumption} (b) shows the memory consumption of tensors across different pipeline stages.
For all LLMs examined in our study, the active tensors account for only less than 14\% (1.7\% on average) of the total GPU memory capacity, although their total memory usage greatly exceeds the GPU memory capacity. 
Most tensors in GPU memory are inactive and can be offloaded to low-cost SSDs, thus, we can best utilize the GPU memory for tensors that will be used by kernels in the near future. 

\begin{figure}[t]
    \centering
    \includegraphics[width=1.0\linewidth]{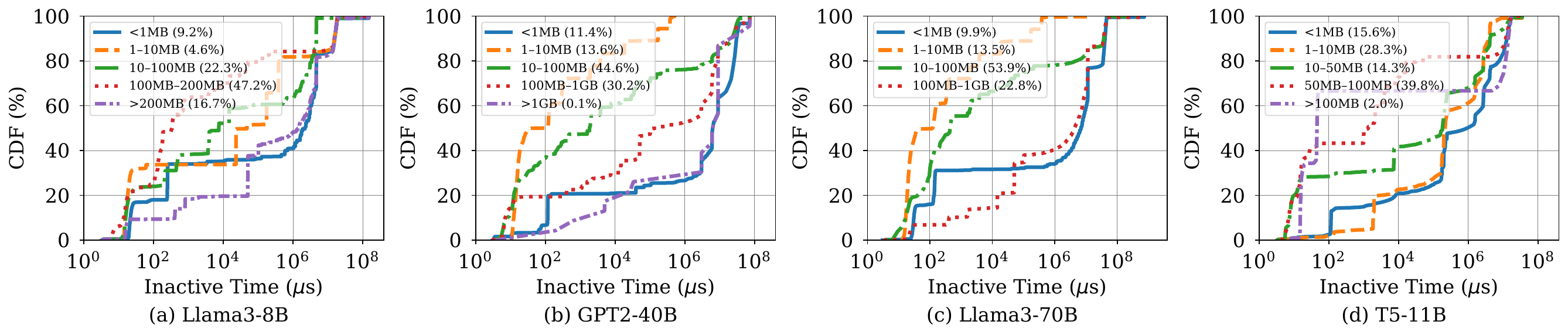}
    \vspace{-4ex}
    \caption{The distribution of inactive periods of tensors. 
    }
    \label{fig:inactive_tensor_distribution}
    \vspace{-3ex}
\end{figure}

\begin{wrapfigure}{r}{0.33\linewidth}
    \vspace{-3ex}
    \includegraphics[width=\linewidth]{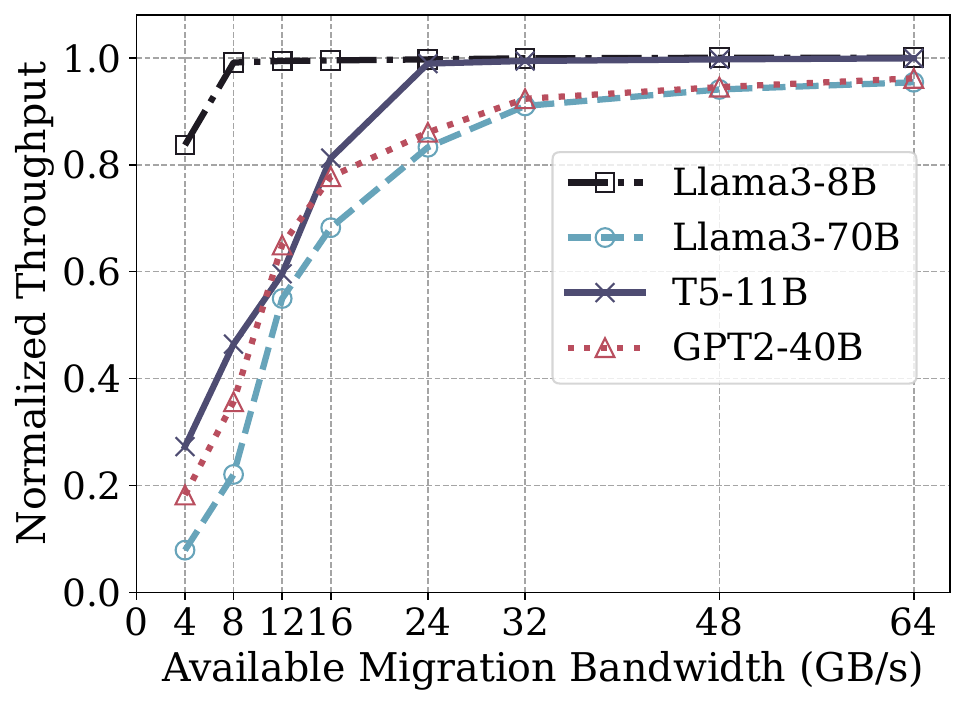}
    \vspace{-4ex}
    \caption{Roofline analysis with different migration bandwidths. The training performance is normalized to the ideal case assuming GPU memory is infinite.}
    \label{fig:roofline_analysis}
    \vspace{-3ex}
\end{wrapfigure}

\noindent
\textbf{Long inactive periods of inactive tensors.}
To understand how long the inactive tensors remain inactive and how much GPU memory they consume, we study the distribution of their inactive periods, as shown in \cref{fig:inactive_tensor_distribution}. 
For all the models we study, most tensors have sizes that range from 10MB to 1GB. 
We observe that more than 40\% of these tensors remain inactive for more than $10^4$ microseconds.
These inactive periods are longer than the time needed to migrate these tensors to SSDs at a bandwidth of 6.5 GB/s. 
With this insight, we can ensure that these tensors can be migrated efficiently without introducing negative impact on the training performance. 


The long inactive periods are the cause of the sparse tensor access pattern and high compute intensity of LLMs.
From a spatial perspective, although LLMs have tens or hundreds of layers, many tensors are used within only a single layer.
From a temporal perspective, the compute-intensive kernels (e.g., attention) in each layer take a considerable amount of time, providing rich opportunities for \name{} to overlap the computation with the migration of inactive tensors.

We also observe that, compared with traditional DNN models, the higher compute intensity and larger model sizes of LLMs lead to substantially longer inactive periods.
For example, in BERT-Large \cite{devlin2018bert}, 48\% of tensors are larger than 100MB, and the inactive periods of more than 60\% of these large tensors are two orders of magnitude shorter than those in LLMs.

\vspace{-1.4ex}
\subsection{Bandwidth Requirement for Tensor Migration}
\label{subsec:bandwidth}

We now study how much migration bandwidth is needed for offloading to achieve near-ideal training performance.
We quantify the roofline performance of different LLM models under different migration bandwidths available to each GPU.
To facilitate this study, we build a performance model to estimate the training time.
In the model, we assume that each kernel's execution time is the same as the value collected in our characterization study.
We simulate tensor migration at the designated bandwidth, and check whether the tensors needed by the kernel are already in GPU memory or not.
If they are still being migrated due to limited SSD bandwidth, the waiting time for the migration is added to the total training time.
\cref{fig:roofline_analysis} shows the normalized roofline training throughput of LLMs under different migration bandwidths.
We observe that a bandwidth of 32 to 48 GB/s is sufficient to achieve near-ideal performance for LLMs.
Such a bandwidth requirement can be easily achieved by aggregating multiple commodity SSDs (e.g., an SSD array), demonstrating the feasibility of \name{}. 
%


\vspace{-2ex}

%% file: design.tex
\section{\name{} Design and Implementation}

\vspace{-2ex}

\begin{figure}[t]
    \centering
    \includegraphics[width=0.93\linewidth]{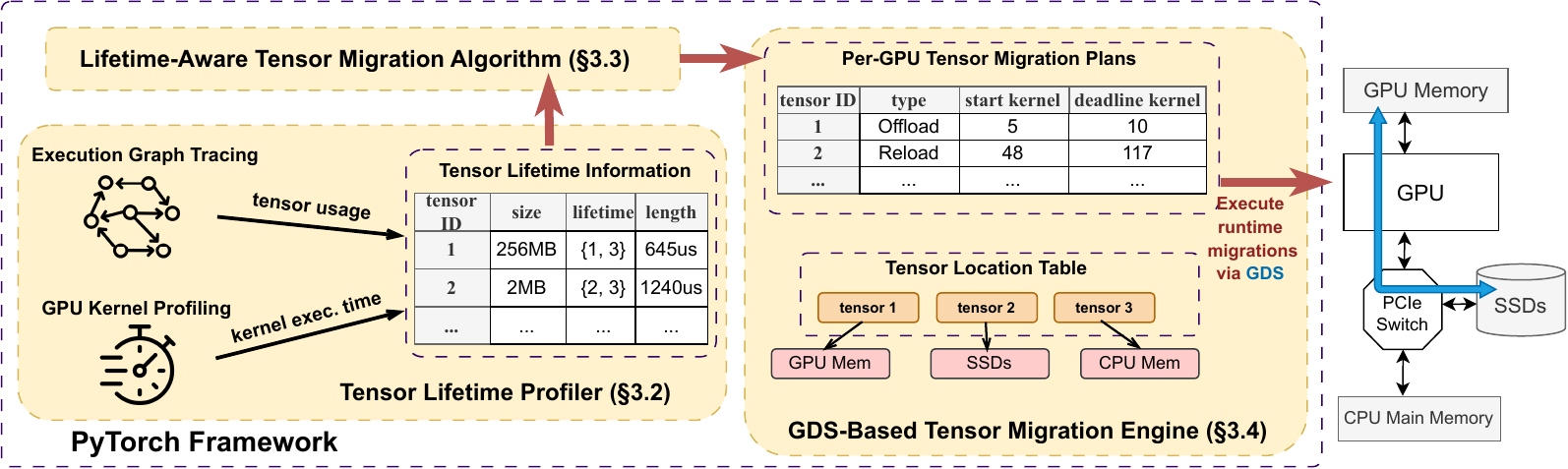}
    \vspace{-1.2ex}
    \caption{System overview of \name{}.}
    \label{fig:overview}
    \vspace{-3ex}
\end{figure}

We show the system overview of \name{} in \cref{fig:overview}.
Given an LLM, \name{}'s tensor lifetime profiler works with PyTorch to track  tensor sizes and lifetimes (\cref{sec:profiling_tool}). In the first few training iterations, \name{} traces the execution graph and collects the execution time of each kernel.
Since the training follows the same execution graph in subsequent iterations, the tensor activity patterns remain the same. The tensor migration algorithm (\cref{sec:algorithm}) creates a tensor migration plan that (1) maximally overlaps computation and migration, and (2) minimizes migration traffic.
The algorithm iteratively selects the best offloading candidates until the required GPU memory fits within the actual GPU memory capacity.
For the migration destination, it prefers to migrate tensors to SSDs.
Once the SSD bandwidth is saturated, it also uses available CPU memory.
During LLM training, \name{}'s tensor migration engine transparently executes the migration plan (\cref{sec:engine}).

\subsection{Tensor Lifetime Profiler}
\label{sec:profiling_tool}


\noindent \textbf{Tracking tensors.}
\name{} instruments PyTorch framework to track tensors and measure kernel execution time at runtime.
A tensor is considered active in one of the following three scenarios.

First, a tensor is active when it is the input or output of a PyTorch CUDA operator. However, instrumenting PyTorch to track every operator is challenging, as there are thousands of operators.
Instead, we leverage PyTorch's automatic operator generator, which produces source code for each operator, 
to insert profiling code that will mark all input and output tensors as active when the operator is executed at runtime.
Second, for tensors that are involved in inter-GPU communication, they should be active in GPU memory.
Third, a tensor is considered active when PyTorch explicitly checks whether it resides in GPU memory.
This happens when updating optimizer states. 
For the second and third scenarios, since there are only a few communication operators and PyTorch checks in total, we directly set the corresponding tensors as active in the source code.

\begin{figure}[!tph]
    \centering
     \vspace{-1.5ex}
    \includegraphics[width=0.8\linewidth]{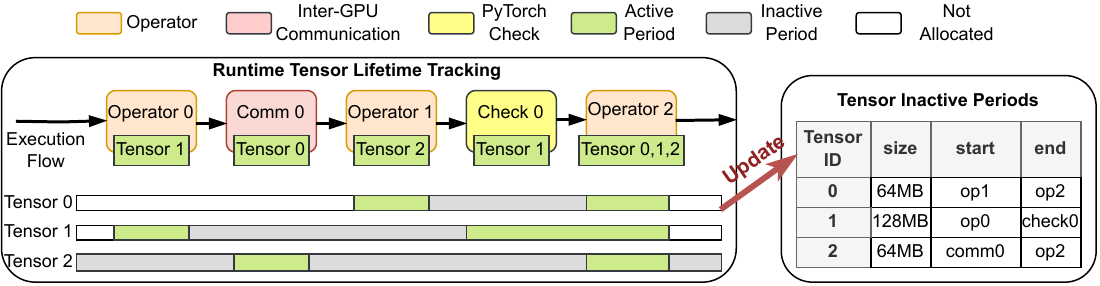}
    \vspace{-1ex}
    \caption{\name{} tracks and analyzes tensor activity patterns for the tensor migration algorithm.}
    \label{fig:tensor_lifetime}
    \vspace{-1ex}
\end{figure}

\noindent \textbf{Analyzing tensor activity patterns.}
\cref{fig:tensor_lifetime} shows how tensor information is collected at runtime.
To understand when a tensor consumes GPU memory and when it must reside in GPU memory, we need to collect its \textit{tensor size and active time}.
When an operator is executed, the instrumented code records the \textit{tensor size} and \textit{active time} for the corresponding tensor. The profiler will calculate the inactive time period based on the duration between its active states. 
Specifically, for intermediate tensors such as gradients and activations (Tensor 0 and Tensor 1 in \cref{fig:tensor_lifetime}) that will be deallocated immediately after its computation completes, we quantify its inactive time period as the time interval between the two active periods. 
For global tensors such as model weights and optimizer states (Tensor 2 in \cref{fig:tensor_lifetime}) that are used across multiple training iterations, they are allocated before training starts and never deallocated during training. Therefore, for some cases, the profiler may need to calculate its inactive period based on the active states across two iterations.  

\vspace{-1.4ex}
\subsection{Lifetime-aware Tensor Migration Algorithm}
\label{sec:algorithm}

The lifetime-aware tensor migration algorithm iteratively finds the best offloading candidates 
in each LLM training iteration, until the required GPU memory is below the capacity limit. 
By tracking the amount of required GPU memory and the storage I/O bandwidth utilization, the algorithm is able to evaluate the potential benefits of tensor offloading. We discuss our key ideas as follows.

\begin{figure}[t]
    \centering
    \includegraphics[width=0.92\linewidth]{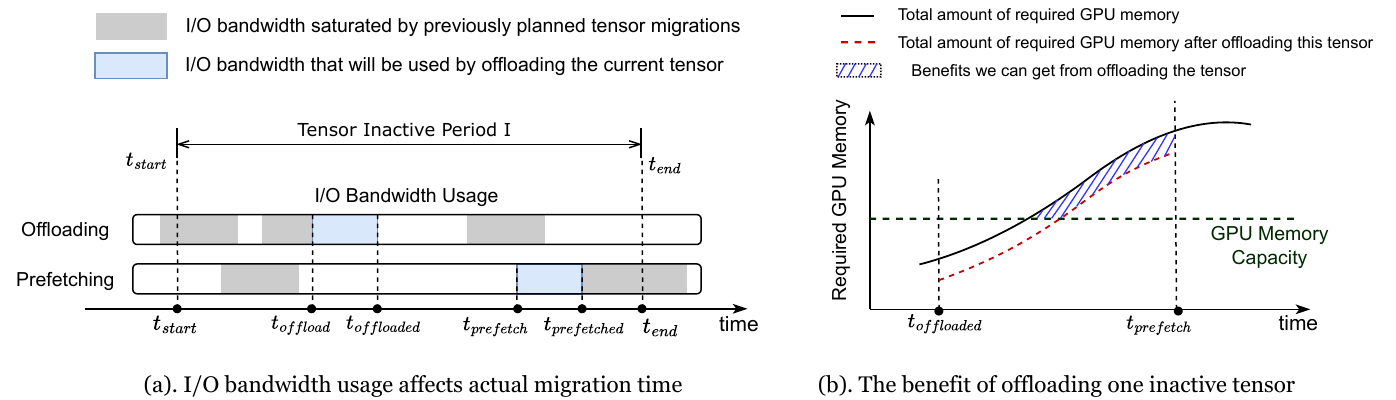}
    \vspace{-1ex}
    \caption{Illustration examples to explain the key insights of lifetime-aware migration algorithm.}
    \vspace{-3.2ex}
    \label{fig:algorithm_ideas}
\end{figure}

\noindent\textbf{Storage I/O bandwidth-aware planning.} 
To offload an inactive tensor, we wish to keep it out of GPU memory as long as possible. Therefore, ideally, we would offload it as soon as it becomes inactive and prefetch it instantly before it is needed by the subsequent kernel.
However, in reality, the time when we can offload and prefetch the inactive tensor is greatly affected by the storage bandwidth usage.
For example, in \cref{fig:algorithm_ideas}(a), the inactive period \textit{I} of the tensor is from $t_{start}$ to $t_{end}$.
However, since the I/O bandwidth is occupied by previously planned migrations, we have to delay the actual offloading until $t_{offload}$ and start the actual prefetching by $t_{prefetch}$.
This means that we can only reduce the memory consumption from $t_{offloaded}$ to $t_{prefetch}$.
When planning tensor migrations, our algorithm tracks the estimated storage I/O bandwidth usage and calculates the reduction in memory consumption in an I/O-aware manner (Line 4-5 and 16-17 in \cref{algo:migration_algorithm}).

\vspace{-1.2ex}
\begin{algorithm}
\scriptsize
\caption{Lifetime-Aware Tensor Migration Planning}
\begin{algorithmic}[1]
\REQUIRE Set of tensor inactive periods $I = \{(t_i, s_i, start_i, end_i)\}$ where $t_i$ is tensor ID, $s_i$ is size, and $[start_i, end_i]$ defines the kernel range of inactivity; GPU memory capacity $M_{GPU}$; Estimation of total amount of required GPU memory $M = [m_0, m_1, ..., m_{N-1}]$ across $N$ kernels; Kernel execution times $T = [\tau_0, \tau_1, ..., \tau_{N-1}]$; I/O bandwidth usage states

\ENSURE Migration plan list $P = \{(t_i, trigger\_time, deadline, target)\}$

\STATE Initialize migration plan list $P = \{\}$ and estimation of required GPU memory $M' = M$
\WHILE{peak memory consumption $\max(M') > M_{GPU}$}
    \FOR{each inactive period $(t, s, start, end) \in I$}
        \STATE Determine earliest feasible offload completion time $t_{offloaded}$ by analyzing available offloading bandwidth
        \STATE Determine latest required prefetch start time $t_{prefetch}$ by analyzing available prefetching bandwidth
        \IF{$t_{offloaded} < t_{prefetch}$}
              \STATE Identify critical memory pressure regions $C = \{k | M'[k] > M_{GPU}, k \in [t_{offloaded}, t_{prefetch}]\}$
            \STATE Calculate the offloading benefit $B = s \times \sum_{k \in C} \tau_k$
        \ENDIF
    \ENDFOR
    \STATE Select inactive period with best offloading cost-benefit $(B/s)_{max}$
    \IF{no viable offloading candidate exists}
        \STATE \textbf{break}
    \ENDIF
    \STATE Add migration plans $(t, start, t_{offloaded}, SSD/CPU)$ and $(t, t_{prefetch}, t_{prefetched}, GPU)$ to $P$
    \STATE Update $M'$ by subtracting tensor size $s$ from affected kernels
    \STATE Update I/O bandwidth usage states
    \STATE Remove selected inactive period from consideration
\ENDWHILE
\RETURN $P$

\end{algorithmic}
\label{algo:migration_algorithm}
\end{algorithm}
\vspace{-1ex}

\noindent\textbf{Quantify the benefit and cost of tensor offloading.}
To fully utilize the available I/O bandwidth, we want to prioritize offloading large tensors with long inactive periods.
Following this principle, our algorithm searches for the best offloading candidate by estimating its benefit and cost.
To quantify the benefit, at a given time \textit{T}, we define \textit{critical memory pressure} as the part of GPU memory consumption that exceeds the capacity.
The benefit of a tensor migration is defined as the integral of the reduction in critical memory pressure over time, as illustrated by the shaded area in \cref{fig:algorithm_ideas}(b).
We quantify the cost as the sum of offloading and prefetching time of tensors.
\name{}'s migration algorithm sorts all candidates by their benefit-to-cost ratio.
It then selects the tensor with the highest ratio for migration in the current iteration. This procedure is shown in Line 6-11 in \cref{algo:migration_algorithm}.

\noindent\textbf{Decide offloading destination.}
\name{} prioritizes SSDs because of their large capacity and low cost.
When the estimated SSD bandwidth is saturated, \name{} will make the best effort to migrate tensors to the host memory.
\name{} allows users to define the maximum amount of host memory that can be used for tensor migration.
Internally, the migration algorithm tracks the estimated host memory consumption.
If it has already reached the user-defined limit, even if SSD bandwidth is saturated, \name{} will not offload tensors to the host memory. 

\noindent\textbf{Minimize kernel stalls.}
If the required GPU memory exceeds the available GPU memory capacity, and \name{} cannot offload tensors to SSDs or host memory, \name{} will stall kernel execution to wait for more inactive tensors to be offloaded.
It will also wait for the tensors needed by the kernel to be migrated back into GPU memory. This would cause storage I/O bandwidth contention. 
Since the next kernel will stall until the needed tensors are migrated to GPU memory, these migrations should complete as early as possible to avoid stalling the GPU training process.
Therefore, \name{} marks these migrations that must finish before the next kernel as `urgent'. At runtime, the tensor migration engine (\cref{sec:engine}) always prioritizes these urgent migrations over other pending migrations.

\vspace{-1ex}
\subsection{Tensor Migration Engine Using GPUDirect Storage}


\label{sec:engine}

To execute the migration plan, we need to locate the addresses of the tensors to be migrated, based on the tensor identifier. \name{} maintains a hashmap-based tensor location table in PyTorch to map tensor identifiers to their current devices and addresses. 

The migration engine migrates tensors between GPU memory and external memory.
For tensor migrations between GPU memory and SSDs, we use GPUDirect storage to achieve direct data transfer between them.
Our choice of GPUDirect storage is motivated by the potential scalability limitations of host CPU in multi-GPU systems.
For example, on an 8-GPU system, to achieve near-ideal performance, each GPU requires 32 to 48 GB/s bidirectional bandwidth (see \cref{sec:study}).
With GPUDirect storage, we can directly connect 8 SSDs to each GPU via PCIe Gen5 switches to meet the bandwidth demand. For the conventional approach that uses the host CPU to first read data from SSDs to the host memory, and then uses \texttt{cudaMemcpy} to move data to the GPU,  the redundant data copy not only causes performance overhead but also wastes precious CPU cycles~\cite{gpudirectstorage,bam:asplos23}.
But as discussed, \name{} still supports migration between GPUs and CPU memory.

\vspace{-1ex}

\vspace{-1.5ex}

%% file: eval.tex
\section{Experiments}

\vspace{-2ex}

We show that
(1) \name outperforms state-of-the-art offloading frameworks by 1.47$\times$ on average when training LLMs that greatly exceed GPU memory capacity (\cref{subsection:training_throughput});
(2) Compared to the case of training LLMs using only GPU memory, \name{} reduces the cost by up to 5.41$\times$ (\cref{subsection:cost_efficiency});
(3) Compared to existing offloading frameworks, \name{} improves the cost efficiency by 1.45$\times$ on average (\cref{subsection:cost_efficiency}); 
(4) \name{} achieves better throughput than ZeRO-Infinity even with less CPU memory and fewer SSDs (\cref{subsection:sensitivity_analysis}).


\vspace{-1ex}
\subsection{Experimental Setup}
\label{subsection:setup}

\noindent \textbf{Models.}
We evaluate \name{} with Llama3-8B, Llama3-70B \cite{paper:llama}, and Granite-code-base-8B \cite{website:granite}.
We use C4 \cite{paper:c4} as our training dataset. 
To study how different memory demands impact the performance of \name{}, we use batch sizes ranging from 16 to 128 and sequence lengths from 1,024 to 8,192.



\begin{wraptable}{r}{0.4\textwidth}
  \centering
  \vspace{-2ex}
  \caption{Our GPU server configuration.}
  \vspace{-1ex}
  \scalebox{0.6}{
  \begin{tabular}{|l|l|}
     \toprule
     \hline
     GPU                      & 2$\times$ NVIDIA H100 NVL                 \\ \hline
     GPU Memory               & 94GB HBM per GPU                   \\ \hline
     CPU                      & 2$\times$ AMD EPYC 9334         \\ \hline
     CPU Memory               & 1.5TB DDR5 (64GB $\times$24)           \\ \hline
     Interconnect             & PCIe Gen5                   \\ \hline
     SSDs                     & 8$\times$ Samsung 990 PRO 2TB          \\ \hline
     SSD Read/Write Bandwidth & 6.7/6.5 GB/s per SSD               \\ \hline
  \bottomrule
  \end{tabular}
  }
  \vspace{-2ex}
  \label{tab:system_setup}
\end{wraptable}

\noindent \textbf{Hardware configuration and ML framework.}
\cref{tab:system_setup} shows the hardware configuration used in our experiments.
Due to the limited PCIe slots on our machine, we can only install at most 8 PCIe SSDs.
When evaluating \name{}, we use 2 H100 GPUs and 2 RAID-0 arrays with 4 SSDs in each array. Each RAID-0 array is logically assigned to one GPU, providing approximately 16 GB/s bandwidth for tensor migrations.
We use PyTorch 2.5.0 \cite{pytorch} and TorchTitan \cite{paper:torchtitan} to train LLMs.


\noindent \textbf{Baselines.} We compare \name{} with the \textit{Ideal case, ZeRO-Offload, and ZeRO-Infinity}. 
\noindent The \textbf{\textit{Ideal case}} assumes that all GPUs in the system have infinite on-board memory, which gives the theoretical best training performance. 
\noindent \textbf{\textit{ZeRO-Offload}} \cite{paper:zero-offload} and \textbf{\textit{ZeRO-Infinity}} \cite{paper:zero-infinity} are popular offloading-based training systems. ZeRO-Offload offloads tensors from GPU memory only to CPU memory, while ZeRO-Infinity leverages both SSDs and CPU memory to expand GPU memory.
\noindent \textbf{\textit{\name{-SSD}}} and \textbf{\textit{\name{-Mixed}}} are two variations of \name{}. \name{-SSD} only migrates tensors to low-cost SSDs, while \name{-Mixed} uses both SSDs and CPU memory.
To make a fair comparison, we let \name{-Mixed} use the same amount of CPU memory for tensor migration as ZeRO-Infinity.

In our evaluation, we aim to compare \name{} with the best performance achievable by ZeRO.
Therefore, for the parallelization strategy, we use tensor parallelism for ZeRO series, as it delivers optimal multi-GPU training performance. 
Moreover, we split each batch into multiple micro-batches in order to amortize the well-known performance bottleneck \cite{paper:smart-infinity} of ZeRO's CPU-based optimizers.
In addition, although activation checkpointing \cite{paper:checkmate, paper:gpipe, paper:kuaishouoffloading, beaumont2021efficient} can reduce memory consumption, we disable it because it degrades ZeRO's training throughput.

In terms of training precision, we use full-precision training in all experiments.
Though mixed-precision training \cite{paper:mixed_precision_training} is popular, the different memory requirements of the mixed-precision training strategies used by TorchTitan and ZeRO will lead to an unfair comparison.
Specifically, when we enable mixed-precision training, in the ZeRO series, all tensors in the GPU are represented by 16-bit floating points, while in TorchTitan, most tensors, including model weights, gradients, and optimizer states still remain in 32-bit floating point format.
Such differences in numerical formats of tensors can lead to significant differences in GPU memory requirements for the same model, resulting in an unfair scenario where \name{} has to migrate larger tensors than ZeRO.

\vspace{-1.4ex}

\subsection{End-to-end Performance}
\label{subsection:training_throughput}
\begin{figure*}[h!t]
    \centering
    \begin{minipage}{\linewidth}
        \centering
        \includegraphics[width=1.0\linewidth]{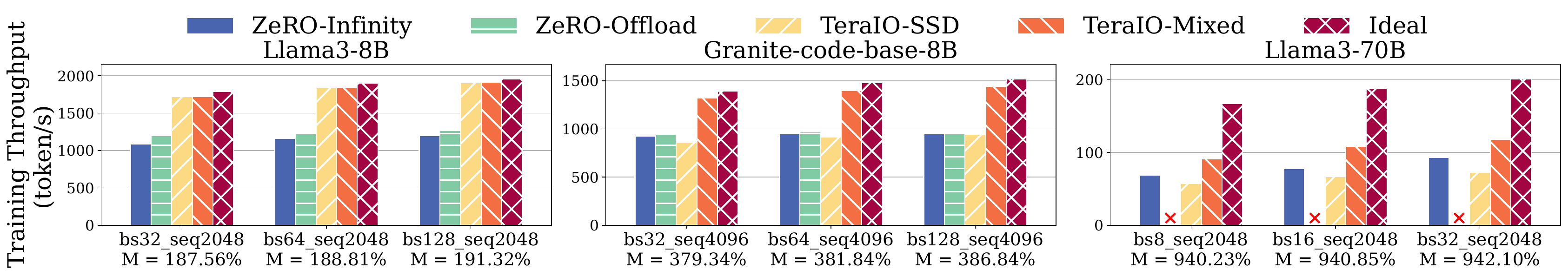}
        \vspace{-3ex}
        \caption*{(1). Average training throughput with different batch sizes.}
        \vspace{0ex}
    \end{minipage}
    \begin{minipage}{\linewidth}
        \centering
        \includegraphics[width=1.0\linewidth]{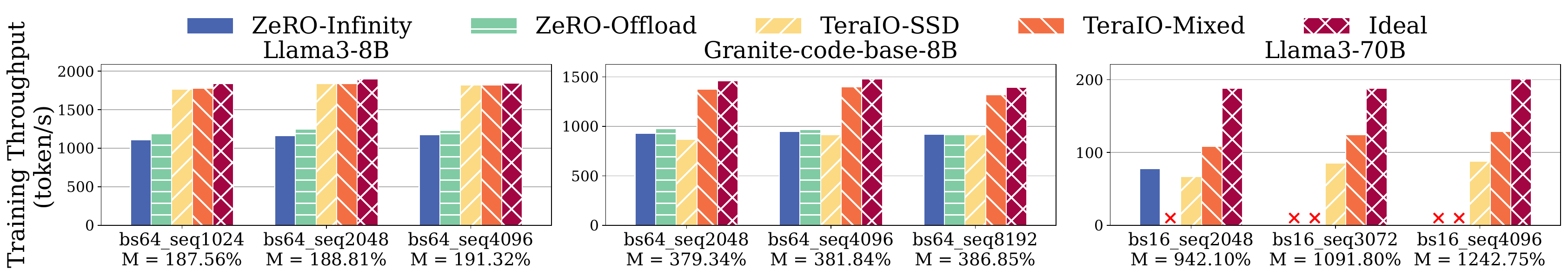}
        \vspace{-3ex}
        \caption*{(2). Average training throughput with different sequence lengths.}
    \end{minipage}
    \vspace{-1ex}
    \caption{Average training throughput of Llama3-8B, Granite-code-base-8B and Llama3-70B with different batch sizes and sequence lengths. \texttt{bs} is the batch size. \texttt{seq} is the sequence length. \texttt{M} is the overall peak memory consumption of the LLM training on one GPU w.r.t. the GPU memory capacity. "\texttt{$\textcolor{red}{\times}$}" means the framework failed to train this model due to out-of-memory errors.}
    \label{fig:training_throughput}
    \vspace{-1ex}
\end{figure*}

We show the end-to-end average training throughput of Llama3-8B, Granite-code-base-8B and Llama3-70B with different batch sizes and sequence lengths in Figure~\ref{fig:training_throughput}. On average, \name{} outperforms ZeRO-Offload and ZeRO-Infinity by 1.47$\times$. Compared to the ideal system assuming unlimited GPU memory, \name{} achieves 80.7\% of the ideal performance.

\vspace{-1.4ex}

\noindent \textbf{Training throughput.}
As shown in Figure~\ref{fig:training_throughput}, when training Llama3-8B and Granite-8B, ZeRO-Offload achieves 65.9\% and 65.5\% of the ideal performance, respectively.
Though ZeRO-Infinity takes extra time to migrate tensors further from CPU memory to SSDs, its training throughput is slightly lower than ZeRO-Offload.
Such a subtle performance difference comes from the high aggregate bandwidth of 8 SSDs and the relatively small sizes of optimizers and parameters offloaded to SSDs.
For Llama-70B, ZeRO-Offload fails because the host memory capacity is too small to store offloaded tensors of such a large model.
Since SSDs offer larger capacity, ZeRO-Infinity can still train Llama3-70B.
However, it only achieves 43.0\% of the ideal performance, because its coarse-grained offloading scheme cannot efficiently utilize the limited SSD bandwidth to migrate larger tensors.


\name{} outperforms the ZeRO series by up to 1.59$\times$.
For Llama3-8B, both \name{-Mixed} and \name{-SSD} can achieve near-ideal performance, demonstrating the effectiveness of our lifetime-aware tensor migration algorithm in choosing the most beneficial tensor to migrate.
Moreover, we find that even though 2 H100 GPUs can provide sufficient GPU memory to train Llama3-8B without memory expansion, \name{} allows us to increase the (micro-)batch size, improving the throughput by 9\%.
For Granite, \name{-SSD} achieves similar performance to the ZeRO series, while \name{-Mixed} can still deliver near-ideal performance by utilizing CPU memory.
The result shows that as the model size increases, achieving near-ideal performance requires not only our lifetime-aware tensor migration algorithm, but also migration bandwidth higher than what 4 SSDs can provide.
For Llama-70B, since its memory requirement significantly exceeds GPU memory capacity, higher migration bandwidth is needed.
Even \name{-Mixed} can only achieve 59.7\% of the ideal performance.
Nonetheless, it still outperforms ZeRO-Infinity by 1.33$\times$.

\noindent \textbf{Impact of varying batch size and sequence length.}
As batch size and sequence length vary, the training throughput of \name{-Mixed} is always the closest to the ideal throughput among all offloading frameworks, while \name{-SSD} delivers similar or better performance than the ZeRO series.
For the same model, increasing the batch size doesn't raise memory requirements, since each batch is split into micro-batches that contain the same number of training samples.
However, when the sequence length increases, the memory requirement also increases.
Therefore, when the sequence length is longer than 3,072, ZeRO-Infinity fails to train the model since the GPU memory capacity is smaller than the activation tensors of the model.
In contrast, \name{} can train all model configurations because it can offload any inactive tensor to save memory.

\begin{figure*}[h!]
    \vspace{-1ex}
    \centering
    \includegraphics[width=1.0\linewidth]{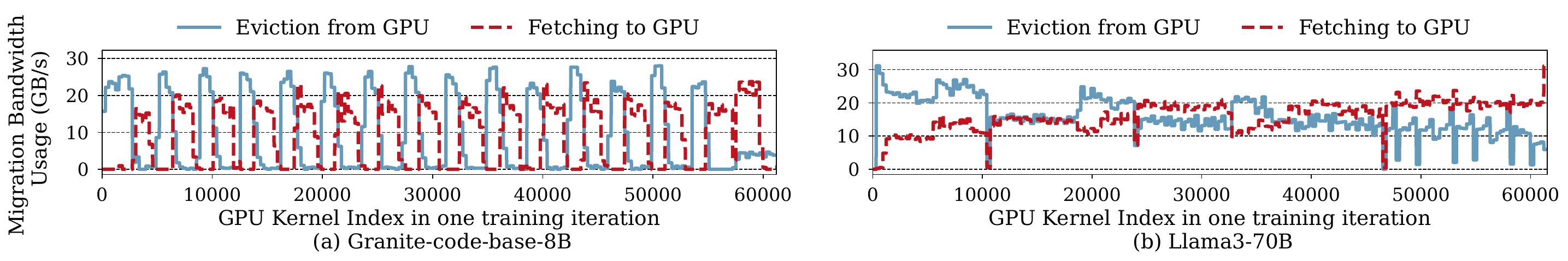}
    \caption{The average migration bandwidth utilization when training the Granite-code-base-8B model and the Llama3-70B model in \name-Mixed. }
    \label{fig:mem_usage_and_migration_activity}
    \vspace{-2ex}
\end{figure*}

\noindent \textbf{Migration bandwidth utilization.}
To further understand why \name{} achieves good performance, we study its migration bandwidth utilization.
Figure~\ref{fig:mem_usage_and_migration_activity} shows that \name{} maintains high utilization of bidirectional migration bandwidth, thanks to our I/O-aware migration algorithm.


\vspace{-1ex}
\subsection{Training Cost}
\label{subsection:cost_efficiency}

\begin{wraptable}{r}{0.77\textwidth}
\tiny
\centering
\vspace{-4ex}
\caption{Cost of each device used by \name{} and other baselines for Llama3-70B training. Prices are quoted from \textit{Exxact} \cite{website:exxact}. The PureGPUs setup contains the minimum number of 8-GPU H100 servers capable for training Llama3-70B without offloading tensors to host memory or SSDs.}
\vspace{0ex}
\scalebox{0.84}{
\begin{tabular}{|c|c|c|c|c|}
\hline
              & \begin{tabular}[c]{@{}c@{}}Server (with 2 H100 GPUs) \\ with 128GB memory\end{tabular} & \begin{tabular}[c]{@{}c@{}}Server (with 2 H100 GPUs) \\ with 1TB memory\end{tabular} & \begin{tabular}[c]{@{}c@{}}$2\times$ Server (with 8 H100 GPUs) \\ with 128GB memory\end{tabular} & \begin{tabular}[c]{@{}c@{}}$8\times$ Samsung 990PRO \\ SSD 2TB\end{tabular} \\ \hline
Cost (\$)     & 84,139.9                                                                                & 91,047.9                                                                              & 499,591.4                                                                                  & 1,360                                                                 \\ \hline
TeraIO-SSD     & \Checkmark                                                                                   &                                                                                      &                                                                                           & \Checkmark                                                                 \\ \hline
TeraIO-Mixed   &                                                                                        & \Checkmark                                                                                 &                                                                                           & \Checkmark                                                                 \\ \hline
ZeRO-Offload  &                                                                                        & \Checkmark                                                                                 &                                                                                           &                                                                      \\ \hline
ZeRO-Infinity &                                                                                        & \Checkmark                                                                                 &                                                                                           & \Checkmark                                                                 \\ \hline
PureGPUs      &                                                                                        &                                                                                      & \Checkmark                                                                                      &                                                                      \\ \hline
\end{tabular}
}
\label{tab:cost}
\vspace{-3ex}
\end{wraptable}




\vspace{-1ex}


To evaluate the cost of \name{}, we summarize the prices of different devices used in each baseline setup in \cref{tab:cost}.
\name{-SSD} needs a server with only 128GB of CPU memory and 8 SSDs since it migrates tensors only to SSDs, while \name{-Mixed} uses both SSDs and 1TB CPU memory for more efficient tensor migration.
Since ZeRO-Offload and ZeRO-Infinity consume a large amount of CPU memory to train LLMs, they both need a server with 1TB of memory.
ZeRO-Infinity additionally uses 8 SSDs in our evaluation.
We also compare with the PureGPU setup, in which all tensors are kept within GPU memory.
To provide enough GPU memory to train Llama3-70B in this setup, we need to pay \$499,591.40 for two servers, each equipped with 8 H100 GPUs.
In comparison, \name{-SSD} and \name{-Mixed} save costs by 5.88$\times$ and 5.41$\times$, respectively.
Compared to ZeRO-Offload and ZeRO-Infinity, since we have similar machine setups, \name{}'s 1.47$\times$ training performance improvement translates into 1.45$\times$ and 1.47$\times$ improved cost efficiency, respectively.




\vspace{-1.4ex}

\subsection{Impact of Varying Number of SSDs and CPU Memory Capacity}
\label{subsection:sensitivity_analysis}

\begin{wrapfigure}{r}{0.55\linewidth}
    \includegraphics[width=\linewidth]{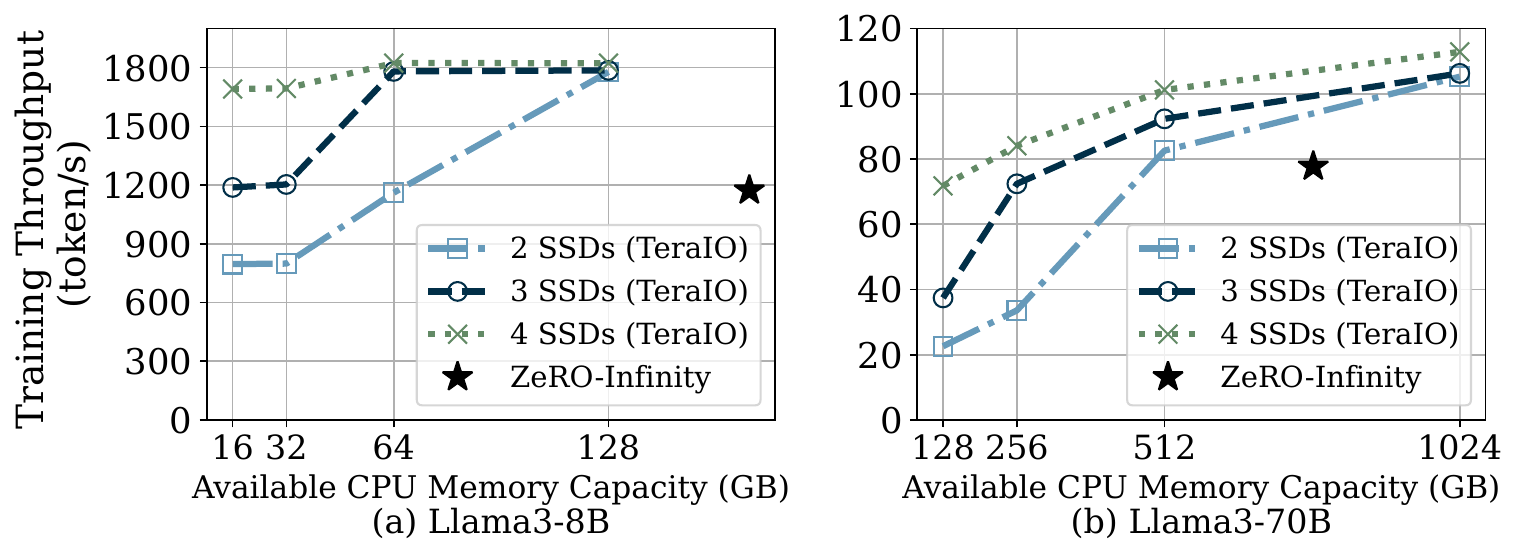}
    \vspace{-3ex}
    \caption{Training throughput as we vary the CPU memory capacity and the number of SSDs per GPU.}
    \label{fig:sensitivity_analysis}
    \vspace{-4ex}
\end{wrapfigure}

Figure~\ref{fig:sensitivity_analysis} shows the training throughput of \name{} as we vary the available CPU memory capacity and the number of SSDs used for each GPU. 
The performance of \name{} scales favorably as we increase the number of SSDs or the CPU memory capacity.
For Llama-8B, \name{} achieves near ideal performance with only 2 SSDs and 64 GB of CPU memory.
Even for Llama-70B, we observe that using only 2 SSDs and 512GB of CPU memory can still achieve 73.1\% of the training throughput obtained with 4 SSDs and 1,024GB of CPU memory.
These results validate our LLM characterization study's key observation: with \textit{lifetime-aware} tensor offloading, we can achieve good performance even with limited hardware resources.

Compared to ZeRO-Infinity, \name{} significantly reduces the CPU memory capacity requirement while achieving better performance.
ZeRO-Infinity requires 170GB and 770GB of CPU memory to train Llama-8B and Llama-70B, respectively, as it has to offload gradients and optimizers into CPU.
In comparison, \name{} achieves better performance even with more limited hardware resources, as it efficiently utilizes both the large capacity of SSDs and extra bandwidth of CPU memory.

\vspace{-2ex}

%% file: ending.tex
\section{Conclusion} 
\label{sec:conclusion}

\vspace{-2ex}

We present \name{}, a lifetime-aware tensor offloading framework that can accurately plan and execute fine-grained tensor offloading and prefetching instructions for LLM training. With predictable tensor activity patterns, \name{} best utilizes precious GPU memory for accelerating GPU training process, while leveraging large-capacity SSDs for lowering training cost. Compared to existing tensor offloading work, \name{} provides a more practical and cost-efficient solution
for LLM training.

%% file: main.bbl
\begin{thebibliography}{10}

\bibitem{paper:flashneuron}
Jonghyun Bae, Jongsung Lee, Yunho Jin, Sam Son, Shine Kim, Hakbeom Jang, Tae~Jun Ham, and Jae~W Lee.
\newblock $\{$FlashNeuron$\}$:$\{$SSD-Enabled$\}$$\{$Large-Batch$\}$ training of very deep neural networks.
\newblock In {\em 19th USENIX Conference on File and Storage Technologies (FAST 21)}, pages 387--401, 2021.

\bibitem{beaumont2021efficient}
Olivier Beaumont, Lionel Eyraud-Dubois, and Alena Shilova.
\newblock Efficient combination of rematerialization and offloading for training dnns.
\newblock {\em Advances in Neural Information Processing Systems}, 34:23844--23857, 2021.

\bibitem{devlin2018bert}
Jacob Devlin, Ming-Wei Chang, Kenton Lee, and Kristina Toutanova.
\newblock Bert: Pre-training of deep bidirectional transformers for language understanding.
\newblock {\em arXiv preprint arXiv:1810.04805}, 2018.

\bibitem{paper:llama}
Abhimanyu Dubey, Abhinav Jauhri, Abhinav Pandey, Abhishek Kadian, Ahmad Al-Dahle, Aiesha Letman, Akhil Mathur, Alan Schelten, Amy Yang, Angela Fan, et~al.
\newblock The llama 3 herd of models.
\newblock {\em arXiv preprint arXiv:2407.21783}, 2024.

\bibitem{website:exxact}
EXXACT.
\newblock Exxact.
\newblock \url{https://www.exxactcorp.com/}, 2025.

\bibitem{website:t5}
Google.
\newblock T5 11b.
\newblock \url{https://huggingface.co/google-t5/t5-11b}, 2025.

\bibitem{gpudirectstorage}
{GPUDirect Storage: A Direct Path Between Storage and GPU Memory}.
\newblock \url{https://developer.nvidia.com/blog/gpudirect-storage/}.

\bibitem{paper:swapadvisor}
Chien-Chin Huang, Gu~Jin, and Jinyang Li.
\newblock Swapadvisor: Pushing deep learning beyond the gpu memory limit via smart swapping.
\newblock In {\em Proceedings of the Twenty-Fifth International Conference on Architectural Support for Programming Languages and Operating Systems}, pages 1341--1355, 2020.

\bibitem{paper:gpipe}
Yanping Huang, Youlong Cheng, Ankur Bapna, Orhan Firat, Dehao Chen, Mia Chen, HyoukJoong Lee, Jiquan Ngiam, Quoc~V Le, Yonghui Wu, et~al.
\newblock Gpipe: Efficient training of giant neural networks using pipeline parallelism.
\newblock {\em Advances in neural information processing systems}, 32, 2019.

\bibitem{website:granite}
IBM.
\newblock Ibm granite.
\newblock \url{https://huggingface.co/ibm-granite}, 2025.

\bibitem{paper:checkmate}
Paras Jain, Ajay Jain, Aniruddha Nrusimha, Amir Gholami, Pieter Abbeel, Joseph Gonzalez, Kurt Keutzer, and Ion Stoica.
\newblock Checkmate: Breaking the memory wall with optimal tensor rematerialization.
\newblock {\em Proceedings of Machine Learning and Systems}, 2:497--511, 2020.

\bibitem{paper:smart-infinity}
Hongsun Jang, Jaeyong Song, Jaewon Jung, Jaeyoung Park, Youngsok Kim, and Jinho Lee.
\newblock Smart-infinity: Fast large language model training using near-storage processing on a real system.
\newblock In {\em 2024 IEEE International Symposium on High-Performance Computer Architecture (HPCA)}, pages 345--360. IEEE, 2024.

\bibitem{paper:deepum}
Jaehoon Jung, Jinpyo Kim, and Jaejin Lee.
\newblock Deepum: Tensor migration and prefetching in unified memory.
\newblock In {\em Proceedings of the 28th ACM International Conference on Architectural Support for Programming Languages and Operating Systems, Volume 2}, pages 207--221, 2023.

\bibitem{paper:scaling_laws}
Jared Kaplan, Sam McCandlish, Tom Henighan, Tom~B Brown, Benjamin Chess, Rewon Child, Scott Gray, Alec Radford, Jeffrey Wu, and Dario Amodei.
\newblock Scaling laws for neural language models.
\newblock {\em arXiv preprint arXiv:2001.08361}, 2020.

\bibitem{paper:memorywall}
Youngeun Kwon and Minsoo Rhu.
\newblock Beyond the memory wall: A case for memory-centric hpc system for deep learning.
\newblock In {\em 2018 51st Annual IEEE/ACM International Symposium on Microarchitecture (MICRO)}, pages 148--161. IEEE, 2018.

\bibitem{paper:tflms}
Tung~D Le, Haruki Imai, Yasushi Negishi, and Kiyokuni Kawachiya.
\newblock Tflms: Large model support in tensorflow by graph rewriting.
\newblock {\em arXiv preprint arXiv:1807.02037}, 2018.

\bibitem{paper:torchtitan}
Wanchao Liang, Tianyu Liu, Less Wright, Will Constable, Andrew Gu, Chien-Chin Huang, Iris Zhang, Wei Feng, Howard Huang, Junjie Wang, et~al.
\newblock Torchtitan: One-stop pytorch native solution for production ready llm pre-training.
\newblock {\em arXiv preprint arXiv:2410.06511}, 2024.

\bibitem{paper:fuyou}
Changyue Liao, Mo~Sun, Zihan Yang, Kaiqi Chen, Binhang Yuan, Fei Wu, and Zeke Wang.
\newblock Adding nvme ssds to enable and accelerate 100b model fine-tuning on a single gpu.
\newblock {\em arXiv preprint arXiv:2403.06504}, 2024.

\bibitem{paper:large_batch_training}
Sam McCandlish, Jared Kaplan, Dario Amodei, and OpenAI~Dota Team.
\newblock An empirical model of large-batch training.
\newblock {\em arXiv preprint arXiv:1812.06162}, 2018.

\bibitem{paper:mixed_precision_training}
Paulius Micikevicius, Sharan Narang, Jonah Alben, Gregory Diamos, Erich Elsen, David Garcia, Boris Ginsburg, Michael Houston, Oleksii Kuchaiev, Ganesh Venkatesh, et~al.
\newblock Mixed precision training.
\newblock {\em arXiv preprint arXiv:1710.03740}, 2017.

\bibitem{paper:pipedream}
Deepak Narayanan, Aaron Harlap, Amar Phanishayee, Vivek Seshadri, Nikhil~R Devanur, Gregory~R Ganger, Phillip~B Gibbons, and Matei Zaharia.
\newblock Pipedream: Generalized pipeline parallelism for dnn training.
\newblock In {\em Proceedings of the 27th ACM symposium on operating systems principles}, pages 1--15, 2019.

\bibitem{paper:angel-ptm}
Xiaonan Nie, Yi~Liu, Fangcheng Fu, Jinbao Xue, Dian Jiao, Xupeng Miao, Yangyu Tao, and Bin Cui.
\newblock Angel-ptm: A scalable and economical large-scale pre-training system in tencent.
\newblock {\em arXiv preprint arXiv:2303.02868}, 2023.

\bibitem{pytorch}
Adam Paszke, Sam Gross, Soumith Chintala, Gregory Chanan, Edward Yang, Zachary DeVito, Zeming Lin, Alban Desmaison, Luca Antiga, and Adam Lerer.
\newblock Automatic differentiation in pytorch.
\newblock In {\em NIPS-W}, 2017.

\bibitem{paper:diffusion_trans}
William Peebles and Saining Xie.
\newblock Scalable diffusion models with transformers.
\newblock In {\em Proceedings of the IEEE/CVF International Conference on Computer Vision}, pages 4195--4205, 2023.

\bibitem{paper:l2l}
Bharadwaj Pudipeddi, Maral Mesmakhosroshahi, Jinwen Xi, and Sujeeth Bharadwaj.
\newblock Training large neural networks with constant memory using a new execution algorithm.
\newblock {\em arXiv preprint arXiv:2002.05645}, 2020.

\bibitem{bam:asplos23}
Zaid Qureshi, Vikram~Sharma Mailthody, Isaac Gelado, Seungwon Min, Amna Masood, Jeongmin Park, Jinjun Xiong, C.~J. Newburn, Dmitri Vainbrand, I-Hsin Chung, Michael Garland, William Dally, and Wen-mei Hwu.
\newblock Gpu-initiated on-demand high-throughput storage access in the bam system architecture.
\newblock In {\em Proceedings of the 28th ACM International Conference on Architectural Support for Programming Languages and Operating Systems, Volume 2}, ASPLOS 2023, page 325–339, New York, NY, USA, 2023. Association for Computing Machinery.

\bibitem{gpt2}
Alec Radford, Jeffrey Wu, Rewon Child, David Luan, Dario Amodei, Ilya Sutskever, et~al.
\newblock Language models are unsupervised multitask learners.
\newblock {\em OpenAI blog}, 1(8):9, 2019.

\bibitem{paper:c4}
Colin Raffel, Noam Shazeer, Adam Roberts, Katherine Lee, Sharan Narang, Michael Matena, Yanqi Zhou, Wei Li, and Peter~J Liu.
\newblock Exploring the limits of transfer learning with a unified text-to-text transformer.
\newblock {\em Journal of machine learning research}, 21(140):1--67, 2020.

\bibitem{paper:zero-infinity}
Samyam Rajbhandari, Olatunji Ruwase, Jeff Rasley, Shaden Smith, and Yuxiong He.
\newblock Zero-infinity: Breaking the gpu memory wall for extreme scale deep learning.
\newblock In {\em Proceedings of the international conference for high performance computing, networking, storage and analysis}, pages 1--14, 2021.

\bibitem{paper:zero-offload}
Jie Ren, Samyam Rajbhandari, Reza~Yazdani Aminabadi, Olatunji Ruwase, Shuangyan Yang, Minjia Zhang, Dong Li, and Yuxiong He.
\newblock $\{$Zero-offload$\}$: Democratizing $\{$billion-scale$\}$ model training.
\newblock In {\em 2021 USENIX Annual Technical Conference (USENIX ATC 21)}, pages 551--564, 2021.

\bibitem{paper:vdnn}
Minsoo Rhu, Natalia Gimelshein, Jason Clemons, Arslan Zulfiqar, and Stephen~W Keckler.
\newblock vdnn: Virtualized deep neural networks for scalable, memory-efficient neural network design.
\newblock In {\em 2016 49th Annual IEEE/ACM International Symposium on Microarchitecture (MICRO)}, pages 1--13. IEEE, 2016.

\bibitem{paper:agent_for_cloud}
Manish Shetty, Yinfang Chen, Gagan Somashekar, Minghua Ma, Yogesh Simmhan, Xuchao Zhang, Jonathan Mace, Dax Vandevoorde, Pedro Las-Casas, Shachee~Mishra Gupta, et~al.
\newblock Building ai agents for autonomous clouds: Challenges and design principles.
\newblock In {\em Proceedings of the 2024 ACM Symposium on Cloud Computing}, pages 99--110, 2024.

\bibitem{paper:stronghold}
Xiaoyang Sun, Wei Wang, Shenghao Qiu, Renyu Yang, Songfang Huang, Jie Xu, and Zheng Wang.
\newblock Stronghold: fast and affordable billion-scale deep learning model training.
\newblock In {\em SC22: International Conference for High Performance Computing, Networking, Storage and Analysis}, pages 1--17. IEEE, 2022.

\bibitem{paper:superneurons}
Linnan Wang, Jinmian Ye, Yiyang Zhao, Wei Wu, Ang Li, Shuaiwen~Leon Song, Zenglin Xu, and Tim Kraska.
\newblock Superneurons: Dynamic gpu memory management for training deep neural networks.
\newblock In {\em Proceedings of the 23rd ACM SIGPLAN symposium on principles and practice of parallel programming}, pages 41--53, 2018.

\bibitem{paper:netllm}
Duo Wu, Xianda Wang, Yaqi Qiao, Zhi Wang, Junchen Jiang, Shuguang Cui, and Fangxin Wang.
\newblock Netllm: Adapting large language models for networking.
\newblock In {\em Proceedings of the ACM SIGCOMM 2024 Conference}, pages 661--678, 2024.

\bibitem{paper:tba}
Kun Wu, Jeongmin~Brian Park, Xiaofan Zhang, Mert Hidayeto{\u{g}}lu, Vikram~Sharma Mailthody, Sitao Huang, Steven~Sam Lumetta, and Wen-mei Hwu.
\newblock Tba: Faster large language model training using ssd-based activation offloading.
\newblock {\em arXiv preprint arXiv:2408.10013}, 2024.

\bibitem{paper:kuaishouoffloading}
Tailing Yuan, Yuliang Liu, Xucheng Ye, Shenglong Zhang, Jianchao Tan, Bin Chen, Chengru Song, and Di~Zhang.
\newblock Accelerating the training of large language models using efficient activation rematerialization and optimal hybrid parallelism.
\newblock In {\em 2024 USENIX Annual Technical Conference (USENIX ATC 24)}, pages 545--561, 2024.

\bibitem{paper:zng}
Jie Zhang and Myoungsoo Jung.
\newblock Zng: Architecting gpu multi-processors with new flash for scalable data analysis.
\newblock In {\em 2020 ACM/IEEE 47th Annual International Symposium on Computer Architecture (ISCA)}, pages 1064--1075. IEEE, 2020.

\bibitem{paper:flashgpu}
Jie Zhang, Miryeong Kwon, Hyojong Kim, Hyesoon Kim, and Myoungsoo Jung.
\newblock Flashgpu: Placing new flash next to gpu cores.
\newblock In {\em Proceedings of the 56th Annual Design Automation Conference 2019}, pages 1--6, 2019.

\end{thebibliography}
